\documentclass[11pt]{article}
\usepackage{amsmath}
\usepackage{amssymb}%

\usepackage{xcolor}
\usepackage{geometry}
\usepackage{mathtools, nccmath}
\usepackage{graphicx}
\usepackage{url} 
\usepackage{lineno}

\usepackage[margin=10pt,font=small,labelfont=bf, labelsep=endash]{caption}
\usepackage{subcaption}

\newcommand\blfootnotef[1]{%
  \begingroup
  \renewcommand\thefootnote{\ssymbol{5}}\footnote{}%
  \addtocounter{footnote}{-1}%
  \endgroup
}

\makeatletter
\newcommand{\ssymbol}[1]{\textsuperscript{\@fnsymbol{#1}}}
\makeatother

\title{Optimizing testing policies for detecting COVID-19 outbreaks}

\author
{Janni Yuval \thanks{All authors contributed equally to this manuscript.} \thanks{MIT.}    \and Mor Nitzan \footnotemark[1] \thanks{The Hebrew University.} 
\and Neta Ravid Tannenbaum \footnotemark[1]  \thanks{Technion}  
\and Boaz Barak \footnotemark[1] \thanks{Harvard University}}

\begin{document}
\maketitle



\begin{abstract}
The COVID-19 pandemic poses challenges for continuing economic activity while reducing health risks. While these challenges can be mitigated through testing, testing budget is often limited. Here we study how institutions, such as nursing homes, should utilize a fixed test budget for early detection of an outbreak. Using an extended network-SEIR model, we show that given a certain budget of tests, it is generally better to test smaller subgroups of the population frequently than to test larger groups but less frequently. The numerical results are consistent with an analytical expression we derive for the size of the outbreak at detection in an exponential spread model. Our work provides a simple guideline for institutions: distribute your total tests over several batches instead of using them all at once. We expect that in the appropriate scenarios, this easy-to-implement policy recommendation will lead to earlier detection and better mitigation of local COVID-19 outbreaks.

\end{abstract}



\section*{Introduction}
\emph{Viral testing} is crucial in monitoring and controlling the spread of Covid-19, and a key factor in policy adjustments~\cite{giordano2020modelling}.
Limitations of financial and medical resources make it imperative to find efficient ways for testing large populations.
Testing can be used to achieve several different goals:
(1) \emph{Diagnostic testing:} Test patients that display symptoms or have had contact with suspected-positive individual, in order to determine the best course of treatment as well as whether isolation is required. 
(2) \emph{Spreading suppression:} Wide-spread testing of asymptomatic individuals in order to detect and isolate infectious individuals, as well as trace their contacts. 
(3) \emph{Outbreak detection:} If the  prevalence of the infection in the population is sufficiently low, randomized testing of asymptomatic individuals can be used for \emph{outbreak detection}. In such a setting the goal of testing is to detect an introduction of an infection into the institution before the outbreak becomes too large.  Once such an outbreak is detected, other emergency measures can be implemented, including contact tracing, large-scale testing, and temporary closures. 

Much of the literature on COVID-19 testing protocols has been focused on \emph{spreading suppression}, where the goal is to test the population frequently enough so that any infected asymptomatic individual will be detected before they had a chance to infect many others.
Combining mass testing with other measures, such as self-isolation and household quarantine could substantially reduce transmission rates \cite{kucharski2020effectiveness}.
In places where the transmission rate is especially large, such as college and university campuses, and in high-risk groups, such as health workers, periodic testing can play a crucial role in reducing transmission \cite{bergstrom2020frequency,evans2020impact,zhang2020periodic,grassly2020comparison}, and many universities already require students and personnel to get periodically tested if they are on campus \cite{booeshaghi2020markedly}. 
Unfortunately, testing the population in such environments less frequently than once in two days is unlikely to reduce the reproductive number below one \cite{paltiel2020assessment, chin2020frequency}, which is the threshold needed to contain an outbreak.

\textcolor{black}{While frequent extensive testing is ideal, not every institution in every locality has the resources to perform it. 
Furthermore, the likelihood of an introduction of infection into an institution depends on both the community prevalence and the size of the institution. If both are sufficiently small so that introductions are relatively rare, and the institution is able to switch to emergency footing to contain and suppress an outbreak once detected, then potentially fewer tests can be used. 
In this work we consider such institutions where new Covid-19 cases are rare and we focus on testing policies for \emph{outbreak detection}.
If the prevalence is high enough or the institution is large enough that introductions of infection are frequent (e.g., daily) \textcolor{black}{or in cases where emergency measures, such as partial or temporary closures, are not an option,} then different suppression strategies need to be used. Such suppression strategies are beyond the scope of this work.}

\textcolor{black}{Outbreak detection is crucial for institutions, such as nursing homes, that are comprised of susceptible populations, but can be useful in other cases as well.
Such institutions have a limited budget of tests and typically do not have ``in house'' testing facilities. Hence in some cases they} rely on mobile testing units that visit the institution at some frequency to test all of its members (see \cite{cms,mobile_test_victoria}).
The question we study is whether there are better ways to distribute the same number of tests to reduce the risk of undetected outbreaks. 

To illustrate this, consider the setting where the budget of tests allows for testing all of the members of the institution once per 28 days.
Two potential policies to distribute the same budget are: (1)  \emph{single batch} policy, where every four weeks there is a testing day in which the entire population is tested, and (2) \emph{frequent-and-partial} policy, where every $x$ days we test a random $x/28$ fraction of the population (for example test a quarter of the population every week).  

Intuitively, there is a trade-off between the two testing policies. If we assume for simplicity that infected individuals do not recover and test results are 100\% accurate, then with the single-batch policy we are guaranteed to detect any outbreak that may exist on each testing day, since the entire population is being tested that day. However, such an outbreak is likely to be quite large by the time it is detected. Instead, with the frequent-and-partial policy, when testing more frequently but only a random subset of the population each time, there is no guarantee of catching an outbreak on a testing day. However, if an infected individual is detected, then the corresponding outbreak might be earlier in its course and therefore of smaller size at the time of detection. 

In this work we show that in many cases, choosing the frequent-and-partial policy, namely distributing the same number of tests into smaller subsets that are performed more frequently, will lead to better outcomes and decrease the expected number of people infected by the time the outbreak is detected.
For example, instead of testing all residents and staff of a nursing home on the same day once a month, it would be better to test twice a month half of the members at each round of testing.
We demonstrate this numerically based on extensive simulations in an elaborate network-based \emph{SEIR model}, that accounts for incubation period, heterogeneity in connections, false negative tests, and testing results lag.
We also show that our results are consistent with the analysis of a simple (and analytically tractable) \emph{exponential spread model}.

\section*{Results}

\subsection*{Outbreak detection - SEIR model}
We start by analyzing the effect of tests distribution on outbreak detection based on a mathematical model for epidemic spread in a population. The SIR (Susceptible-Infectious-Recovered) epidemic model \cite{kermack1927contribution} and its variations such as SEIR (e.g. \cite{allen2008introduction,dottori2015sir}, where \textcolor{black}{the} E stands for Exposed) were recently shown to be useful as a modeling framework for COVID-19 analysis \cite{kissler2020projecting,yang2020modified}.
In such a model, each member of the population may be at a given time either susceptible (S), exposed (E), infectious (I), or recovered (R). In this basic model, a susceptible member of the population can transition to an exposed state by interacting with a member in an infectious state. Once a member is exposed, they will transition to the infectious and then to recovered state. 
We use a network-based SEIR model \cite{seirsplus}, in which individuals can interact via links in a social network with a predetermined structure.
A detailed description of the model along with its parameter values are elaborated in the Methods. 

To study the effect of the distribution of a given budget of tests on outbreak detection we assume a budget of $N$ tests per a period of $T$ days, where $N$ is the population size. Two extreme testing policies are: (1) ``one-batch policy'': test all $N$ individuals together, once every $T$ days, and (2) ``frequent daily policy'': testing {$1/T$} fraction of the population daily. We evaluate the testing policies by their associated ``cost'', which is defined as the number of individuals that are infected at the time the first infectious individual is detected.
In more detail, to study intermediate variations between the 1-batch and frequent daily policies, and to obtain concrete estimates for the number of people infected when the first infectious person is detected under different conditions, we perform simulations with varying testing frequencies (testing once every 28,14 and 1 days) and different values of reproductive number $R$, while holding $T$ (which is proportional to the testing budget) constant at $28$ days. 
\textcolor{black}{When using a frequent testing policies, we divide randomly the population to fixed prearranged subgroups which are tested consecutively}.
We use a fixed population size ($N=500$), \textcolor{black}{where the possibility of interactions between individuals is set by a social network structure, while we also enable random infrequent interactions between any two individuals (Methods).}



\textcolor{black}{We assume that individuals within our population (representing for example individuals within a nursing home) can be infected by an external source (for example, by interaction with a visitor). We initially investigate the epidemic spread in simulations where infections by external sources are infrequent. Specifically, we first assume that there is only a single infection due to an external source at a random time, and later show that the results hold when external sources are more frequent.}
As mentioned above, we expect outbreak detection to be a suitable policy when infections by external sources are infrequent.
Due to the stochastic nature of the simulations we estimate the mean number of individuals that are infected by the first detection of an infectious individual for each testing policy and $R$ value using $400$ realizations.

Since our focus is on outbreak detection, we stop the simulation at the first reported positive test result (accounting for false negatives and reporting lag time).
Any mitigation or suppression efforts following detection will vary  greatly according to context and are beyond the scope of this paper.
The goal of the outbreak detection strategy is to minimize the overall number of individuals that were infected up until the point of detection.

\begin{figure}[htbp]
    \centering
 \includegraphics[width=.95\textwidth]{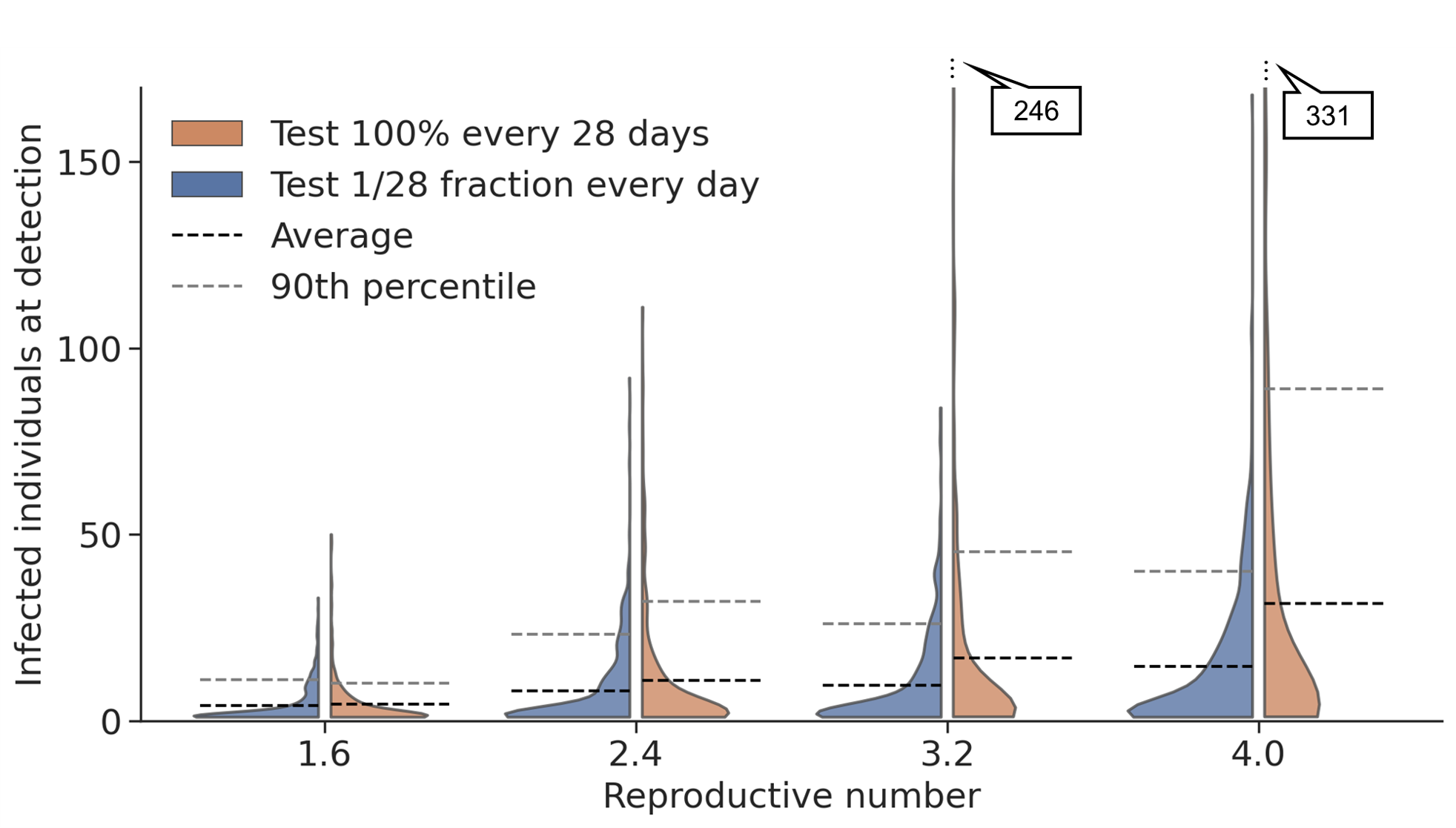}     
    \caption{ \textbf{The number of people infected when the first infectious individual is detected (i.e., cost) for different testing policies and reproductive numbers.}
    \textcolor{black}{
    The distributions of the cost of testing the whole population simultaneously every 28 days (``1-batch policy''; orange) and testing $1/28$ of the population daily (``frequent daily policy''; blue) as a function of the reproductive number. The dotted black lines show the mean cost and the dotted gray lines show the $90^{th}$ percentile of cost for each policy and reproductive number. Both policies have the same testing budget of testing $100\%$ of the population every 28 days. In all simulations used in this figure we consider the introduction of a single \textcolor{black}{exposed} individual at a random time.}
    }
    \label{fig:infectionpolicies}
\end{figure}
\begin{figure}[tbhp]
\centering
   \includegraphics[width=1\linewidth]{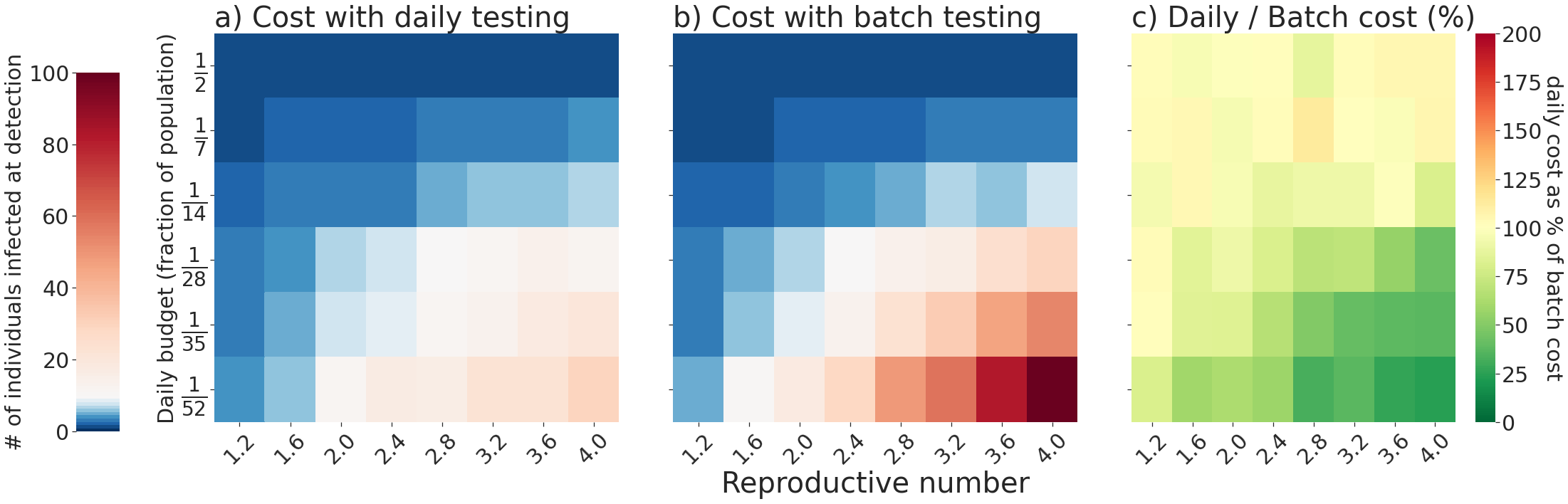}
\caption{\textbf{The mean number of people infected when the first infectious individual is detected (i.e., cost) as a function of the testing budget and reproductive number for different testing policies.}
The mean cost for (a) the ``frequent daily policy'' (test $1/T$ fraction of the population per day) and (b) the ``1-batch policy'' (test 100\% of the population every $T$ days) as a function of the testing budget ($1/T$, where $T$ is the number of days until the whole population is tested once) and the reproductive number.  (c) The ratio between the two policies in percentages ($<100\%$ means that frequent daily policy is better than the 1-batch policy). In all simulations used in this figure we consider the introduction of a single \textcolor{black}{exposed} individual at a random time. }
\label{fig:compare_budget_baseline_notest}
\end{figure}
\begin{figure}[tbhp]
\centering
   \includegraphics[width=1\linewidth]{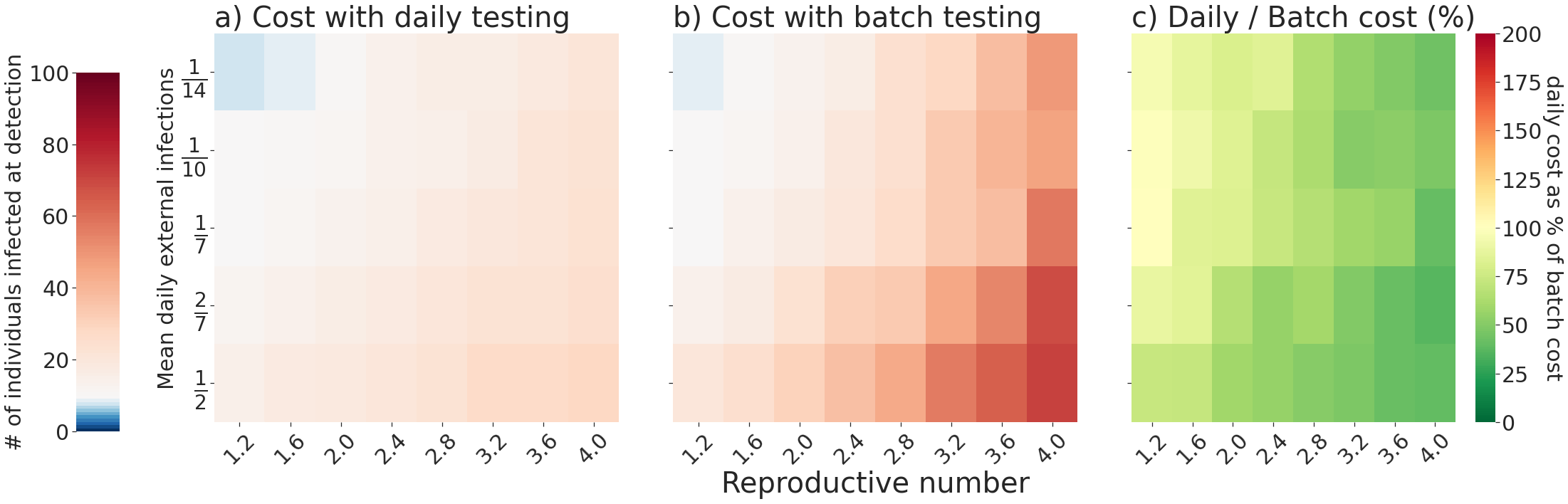}
\caption{\textbf{The mean number of people infected when the first infectious individual is detected (i.e., cost) as a function of the   mean daily number of individuals that are infected from an external source and the reproductive number for different testing policies.}
The mean cost for (a) the ``frequent daily policy'' (test $1/T$ fraction of the population per day) and (b) the ``1-batch policy'' (test 100\% of the population every $T$ days).  (c) The ratio between the two policies in percentages ($<100\%$ means that frequent daily policy is better than the 1-batch policy). \textcolor{black}{Simulations with external infection sources start with zero infected individuals, and in every day there is a probability ($\pi_{\rm{ext}}$) for each susceptible member of the population to get infected (Methods). 
The mean daily external infection is $N\pi_{\rm{ext}}$ where $N$ is the population size. 
The first testing day in the simulation with the 1-batch policy is chosen randomly to be any value between 1 to $T$ days.}
} 
\label{fig:compare_freq_baseline_notest}
\end{figure}

We find that, across varying parameters, the cost is largest for the 1-batch policy (testing the whole population every 28 days), smaller for the 2-batch policy (testing half of the population every 14 days; \textcolor{black}{Fig.~S1}) and smallest for the frequent daily policy (Fig.~\ref{fig:infectionpolicies}). 
Depending on the reproductive number, even a relatively low budget of tests (\textcolor{black}{allowing to test the whole institution once per month}) can still enable detection of an outbreak before it becomes very large, but using the same budget with larger frequency of testing can substantially decrease the size of the outbreak at detection (Fig.~\ref{fig:infectionpolicies}).
Beyond reducing the mean cost, increased testing frequency reduces the probability of missing a large outbreak (\textcolor{black}{Fig.~\ref{fig:infectionpolicies}}).
Our results are robust to changes in the overall testing budget (Fig.~\ref{fig:compare_budget_baseline_notest}), changes in the frequency of individuals who are infected from an external source (Fig.~\ref{fig:compare_freq_baseline_notest}), and changes in the social network structure (Fig.~S2).
Generally, we find that frequent testing is substantially better than the ``1-batch policy'' in cases where an outbreak is expected to be large. For example, in cases where the reproductive number is large (Fig.~\ref{fig:infectionpolicies}), the budget is small (Fig.~\ref{fig:compare_budget_baseline_notest}),  or when there are individuals who are infected from an external source frequently (Fig.~\ref{fig:compare_freq_baseline_notest}), the frequent-testing policy outperforms the 1-batch policy.

\subsection*{Outbreak detection - exponential model}
To support the generality of our results, and to get more intuitive understanding of our findings, we consider a simplified variation of the network-SEIR model which is analytically tractable.
Specifically, we consider an exponential infection spread model (branching process model with an unbounded total population) \cite{anderson1992infectious,kucharski2020early}.
From an epidemiological perspective, this is a model for infection where each infected individual infects exactly $g$ other individuals per time unit $t$. Infected individuals never recover, and testing is 100\% sensitive and specific for infected individuals.
Here we compare two testing policies: "1- batch policy" - test all individuals every $T$ days, and "frequent policy" - test $\epsilon$ fraction of the population every $\epsilon T$ days. 
We define the cost as the overall number of infected individuals at detection, which can be expressed using a single variable - the infection growth rate per testing period ($G \equiv g^T$).
A detailed description of the exponential model and the derivation of the cost for each testing policy appear in the Methods section. 

In this model we prove that:

\begin{itemize}
    \item The expected cost (i.e., overall number of infected individuals) for the ``1-batch policy'' in which we test 100\% of the population every time unit is $\frac{G-1}{\ln G}$ individuals.
    
    \item The expected cost for the ``frequent policy'' where we test $\epsilon$ fraction of the population every $\epsilon$ units equals (in the limit of $\epsilon \rightarrow 0$)  $1+\ln G$ individuals.
\end{itemize}

\begin{figure}[htbp]
    \centering
    \includegraphics[width=4in]{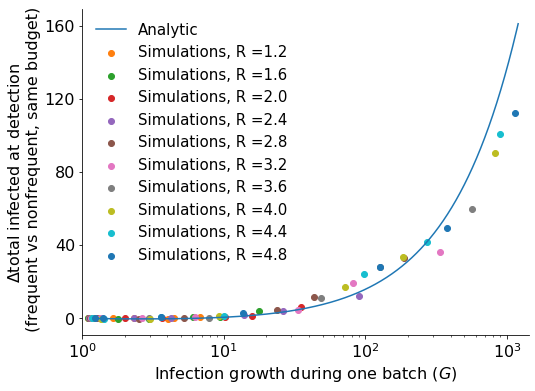}
    \caption{\textcolor{black}{\textbf{The difference in the mean number of infected people when infection is detected (i.e., the cost) between frequent and non-frequent testing as a function of the infection growth rate per testing period.}
    Analytical prediction (blue line) and numerical simulations (dots) of the difference between the mean cost in the ``1-batch policy'' (test 100\% of population every $T$ days) and ``frequent policy'' (test $1/T$ fraction every day in simulations, and testing frequency goes to infinity in the analytical expression), as a function of the infection growth rate per testing period ($G$). A positive  value means that the ``1-batch policy'' has a larger cost than the ``frequent policy''.
    Different colors of dots correspond to different reproductive numbers ($R$), and dots with same colors correspond to different $T$ values ($T =  1,  2,  7, 14, 28, 35$ and $52$ days).
    Results from simulations are obtained by calculating the average over $400$ realizations.}
}
    \label{fig:introfig}
\end{figure}

The difference between the cost of the ``1-batch policy'' and the ``frequent policy'' is positive when $G>3.01$, and the difference grows as $G$ increases (Fig.~\ref{fig:introfig}).
For low $G$ values, the cost of the 1-batch policy can be lower than the cost of the ``frequent policy'', but this difference is never less than $3-e \approx 0.28$ individuals (when $G=e$).  Conversely, the cost of the frequent policy can sometimes be much lower than that of the ``1-batch policy'' (Fig.~\ref{fig:introfig}), especially for low testing budgets (small testing budget increases $G$).

To test the relevance of the analytical expressions derived with the exponential infection spread model to a more realistic scenario, represented here by the extended network-SEIR model, we conduct 
simulations with different $R$ and $T$ values for the  ``1-batch policy'' and the ``frequent daily policy'' where a single external infection is introduced at a random time in the simulation, and $N=2000$. We choose a larger population size for these simulations to avoid situations where a considerable fraction of the population is infected, since in that case we do not expect the analytical results to hold (and outbreak detection policy is probably not very useful).
To enable a comparison between the analytical result and the results from the network-SEIR simulations we need to evaluate $G$ in these simulations. To evaluate $G$ for each combination of $T$ and $R$ (overall 70 combinations), we run $400$ realizations, where each realization is run for $T$ days and is initialized with a single external infection at the initial time step of the simulation (and no testing policy is used).

We find that the cost difference between the ``1-batch policy'' and the ``frequent policy'' derived from the exponential spread model is in close agreement with the cost difference between the ``1-batch policy'' and the ``frequent daily policy'' of the more elaborate network-SEIR model (Fig.~\ref{fig:introfig}). Specifically, the ``frequent testing'' policy performs much better than the ``1-batch policy'' for large $G$ values ($G>10$), and for smaller $G$ values the two policies lead to comparable results (Fig.~\ref{fig:introfig}). 
The difference in the cost between the two testing policies increases when $G$, $R$ or $T$ increase, and this can be explained by:
(1) the cost of ``1-batch policy'' increases roughly linearly for large values of $G$, while the cost of the ``frequent policy'' increases logarithmically with $G$ (these scaling relation are also obtained using network-SEIR model, though the logarithmic growth of the ``frequent daily policy'' has a larger slope than what was derived for the analytical expression, possibly due to the use of infinite testing frequency in analytical derivation; Fig.~S3), and
(2) $G$ increases when $R$ or $T$ increase. Specifically, we show that in the network-SEIR model $G \approx c^{RT}$ (Fig.~S4) where $c$ is a constant that depends on other parameters, such as social network structure, infection period etc.




\section*{Discussion}
In this work we primarily focus on testing policies for outbreak detection in small institutions. We find that given a fixed testing budget, frequent testing of small subgroups within institutions reduces the risk of large undetected outbreaks compared to testing larger groups, but less frequently.
\textcolor{black}{Moreover, as the testing budget decreases and the reproductive number increases frequent testing of small subgroups becomes substantially more effective in limiting the size of outbreaks compared to testing larger groups less frequently.}
Overall, our work suggests frequent testing of subgroups as a strategy to enable safer re-opening of institutions.
We show this both via numerical studies of an elaborate network-based SEIR model, and analytically for the exponential spread model.
Using the exponential spread model we derive an analytic formula for the number of infected people when the outbreak is detected (i.e., the cost) for different testing policies. We show that the cost of frequent sub-population testing grows \emph{logarithmically} with the growth-factor of the infection, as opposed to nearly \emph{linearly} for periodic full-population testing. We demonstrate that these results hold also when using network-based SEIR model. Furthermore, we demonstrate that the frequent subgroup testing is robustly superior to the periodic full-population testing for different testing budgets, reproductive numbers and different rates of external infection sources. 


The parameters we use, including the length of incubation period, infectious period, and the false negative rates, are taken from the literature, and could be refined with further knowledge.
However, we do not expect these to change any of the qualitative results.
We remark that a complete Covid-19 mitigation strategy involves many factors beyond testing, including re-organization, screening for symptoms, and changing testing frequencies based on business role or risk factors.
We do not model such strategies in this paper, beyond their effect on the  reproductive number.
\textcolor{black}{We stress that the aim of this study is to compare
different strategies of randomized testing of asymptomatic individuals \textcolor{black}{constrained by limited budget}, whose goal is to detect an outbreak relatively early (before many infected people are detected following symptoms \textcolor{black}{\cite{gorges2020staffing,mcmichael2020epidemiology}}).}
\textcolor{black}{Future work could test how to effectively combine the testing strategy suggested in this study with different types of tests, such as group tests \cite{dorfman1943detection,shental2020efficient,augenblick2020group} or less sensitive (but cheaper) tests \cite{mina2020rethinking,larremore2020test}, which could reduce the budget necessary for outbreak detection even further.}



While detailed knowledge of institution members roles, connections within the institution, preexisting conditions, and other characteristics could be leveraged  to optimize a testing protocol, in this work we aim to offer a simple and general policy, and therefore consider testing schedules that are independent of individual characteristics. 
Such testing policies, which are  oblivious to symptoms, \textcolor{black}{are} crucial in the discovery of asymptomatic or pre-symptomatic individuals. Overall, our main result is that frequent testing of subpopulations can enhance early outbreak detection of Covid-19  under limited medical and financial resources.

\section*{Methods}

\subsection*{Extended network-based SEIR simulations}

The equations describing the network-SEIR model we use can be expressed as:
\begin{equation}
    P^i(S\rightarrow E) = 
    \left[\pi_{\rm{ext}} + (1-\pi_{\rm{ext}})\left(p \frac{\beta I}{N} + (1-p) \frac{\sum_{j\in C_G(i)} \delta^{ij} \beta \mathbf{1}_{X^j = I}}{|C_G(i)|} \right)\right]\mathbf{1}_{X^=S},
\end{equation}
\begin{equation}
P^i(E\rightarrow I) =  \sigma \mathbf{1}_{X^=E},
\end{equation}
\begin{equation}
P^i(I\rightarrow R) = \gamma {1}_{X^=I},
\end{equation}
where $P^i(X\rightarrow Y)$ is the probability of patient $i$ to move from state X to state Y; $S,E,I$ and $R$ are the susceptible, exposed, infectious and recovered states, respectively; 
$\sigma$ is the rate of progression to infectiousness state (inverse of latent period) and is set to  $1/5.2$ \cite{li2020early};
$\gamma$ is the rate of progression to recovered state (inverse of infectious period) and is set to  $1/6.5$ days$^{-1}$ \cite{yang2020modified};
$\beta$ is the transmissibility of infected individuals where in each simulation $\beta$ is determined by the following formula $\beta = R/\gamma$, and we report the $R$ values instead of $\beta$ values for easier interpretation; 
$C_{G}(i)$ represents close contacts of an individual i, where close contacts are determined by the \textcolor{black}{random network described by Erd{\H{o}}s-R\'{e}nyi random graph \cite{erdHos1960evolution} with a probability of each edge being $15/N$ (average degree $15(N-1)/N \approx 15$)};
$p$ is the proportion of interactions out of the network ($p = \textcolor{black}{}{0.2}$);
$N$ is the population size;
$\mathbf{1}_{Xi = Z}$ is an indicator function where it equals to 1 if the state of node $i$ is $Z$, and equals to $0$ otherwise;
$\pi_{\rm{ext}}$ is the probability to get infected from an external population (outside the population of interest), \textcolor{black}{and we report the overall mean daily external infections in Fig.~\ref{fig:compare_freq_baseline_notest} which is $N\pi_{\rm{ext}}$. The external infection probability is set to be zero in all simulations without external infection sources.}


We assume that some tests can be false negative, and for simplicity we assume that the false negative rate for the exposed population is $100\%$ and for infectious population the false negative rate is $25\%$. These false negative rates are in the range that was found in a previous study \cite{kucirka2020variation}. Furthermore, we assume that there is a lag of one day between taking a test and getting the test result.

\subsection*{Exponential dynamics}
In the \emph{exponential infection dynamics}, we consider one initial infection at time $t_0$. The number of infected people at time $t$ can be written as
\begin{equation}
    I(t)={g}^{t-t_0}
\end{equation}
where $g$ is the growth rate per time unit. 
The testing budget contains $N$ tests for a time period $T$, where $N$  is the  population size. 
 For convenience, we define the growth rate per testing period: $G \equiv g^{T}$.
We define the cost of detecting the outbreak at time $t$ as the number of infected individuals at this time, and we compare between the costs of two policies:
\begin{enumerate}
    \item One batch policy: testing  the entire population every time period $T$.
    \item k batches policy: testing $\frac{1}{k}$ fraction of the population every time period  $\frac{{T}}{k}$.
\end{enumerate}
For the \textit{one batch policy}, the cost is simply $g^{T-t_0}$. Averaging over different values of $t_0$ we get the mean cost of the \textit{one batch policy}:

\begin{equation}
    C_{1}(G)= {\frac{1}{T}}\int_0^{{T}}
   {g}^{T-t_0} dt_0 =\frac{G-1}{{\rm ln}\left({G}\right)}. 
\end{equation}


For the \textit{ k batches policy}, {(and in the limit of $1/k \ll 1$)} the probability that we fail to detect an outbreak at time $t_0+\frac{iT}{k}$ is $(1-\frac{1}{k})^{g^{\frac{iT}{k}}} = (1-\frac{1}{k})^{G^{\frac{i}{k}}}$.
The probability that we detect an outbreak at time $t_0 +\frac{T\ell}{k} $ is equal to the probability we failed to detect the outbreak up until this point (which is $\prod_{i=0}^{\ell -1} (1-\frac{1}{k})^{G^{\frac{i}{k}}} = (1-\frac{1}{k})^{\sum_{i=0}^{\ell -1} G^{\frac{i}{k}}}$) multiplied by the probability to detect the outbreak at time $\frac{T \ell} {k}$ (which is equal to $\left( 1-(1-\frac{1}{k})^{G^{\frac{\ell} {k}}} \right)$).  Since the number of infected people at time $t_0+\frac{T n} {k}$ is $G^{\frac{n} {k}}$,  the cost of the $k$ batches policy can be written as:
\begin{equation}
    C_k(G) = \sum_{n=0}^{\infty} (1-\frac{1}{k})^{\frac{G^{n/k}-1}{G^{1/k}-1}} \cdot \left( 1-(1-\frac{1}{k})^{G^{n/k}} \right) \cdot G^{\frac{n}{k}}.
    \label{eq: C_k sum}
\end{equation}
%
%
%
For $k \rightarrow \infty$,  the summation in equation (\ref{eq: C_k sum}) can be approximated by the following integral:
\begin{equation}
    C_{\infty}(G) \rightarrow \int_{t=0}^{\infty}
    {\rm d}t 
    G^{2t}
    {\rm exp}
    \left(\frac{1-G^t}{{\rm ln}(G)}\right)
    = {\rm ln}(G)+1,
\end{equation}
where  $t=\frac{n}{k}$




\subsection*{Acknowledgements}
We thank Alison Hill for thorough and thoughtful feedback on the manuscript.
BB is supported by NSF awards CCF 1565264 and CNS 1618026 and a Simons Investigator Fellowship.
\textcolor{black}{MN is supported by an Early Career Faculty Fellowship by the Azrieli Foundation.}
JY is supported by the EAPS Houghton-Lorenz postdoctoral fellowship.

\subsection*{Code availability}
Code for the \emph{SEIR dynamics} is available in the GitHub repository \url{https://github.com/boazbk/seirsplus}. 

\subsection*{Data availability}
The data that support the findings of this study are openly available in the GitHub repository \url{https://github.com/boazbk/testingstrategies}.

\subsection*{Competing interests}
The authors declare no competing interests

\subsection*{Supplementary materials}
Figs. S1 to S4 \\


%

\bibliography{pnas-sample}

\newpage




\end{document}